\documentstyle[epsfig]{mn}

\newif\ifAMStwofonts
\ifoldfss
  \ifCUPmtlplainloaded \else
    \NewTextAlphabet{textbfit} {cmbxti10} {}
    \NewTextAlphabet{textbfss} {cmssbx10} {}
    \NewMathAlphabet{mathbfit} {cmbxti10} {} % for math mode
    \NewMathAlphabet{mathbfss} {cmssbx10} {} %  "   "    "
  \fi
  \ifAMStwofonts
    \ifCUPmtlplainloaded \else
      \NewSymbolFont{upmath} {eurm10}
      \NewSymbolFont{AMSa} {msam10}
      \NewMathSymbol{\upi}     {0}{upmath}{19}
      \NewMathSymbol{\umu}     {0}{upmath}{16}
      \NewMathSymbol{\upartial}{0}{upmath}{40}
      \NewMathSymbol{\leqslant}{3}{AMSa}{36}
      \NewMathSymbol{\geqslant}{3}{AMSa}{3E}

    \fi
  \fi
\fi % End of OFSS

\ifnfssone
  \newmathalphabet{\mathit}
  \addtoversion{normal}{\mathit}{cmr}{m}{it}
  \addtoversion{bold}{\mathit}{cmr}{bx}{it}
  \newmathalphabet{\mathbfit} % math mode version of \textbfit{..}
  \addtoversion{normal}{\mathbfit}{cmr}{bx}{it}
  \addtoversion{bold}{\mathbfit}{cmr}{bx}{it}
  \newmathalphabet{\mathbfss} % math mode version of \textbfss{..}
  \addtoversion{normal}{\mathbfss}{cmss}{bx}{n}
  \addtoversion{bold}{\mathbfss}{cmss}{bx}{n}
  \ifAMStwofonts
    \ifCUPmtlplainloaded \else
      \UseAMStwoboldmath
      \makeatletter
      \new@mathgroup\upmath@group
      \define@mathgroup\mv@normal\upmath@group{eur}{m}{n}
      \define@mathgroup\mv@bold\upmath@group{eur}{b}{n}
      \edef\UPM{\hexnumber\upmath@group}
      \new@mathgroup\amsa@group
      \define@mathgroup\mv@normal\amsa@group{msa}{m}{n}
      \define@mathgroup\mv@bold\amsa@group{msa}{m}{n}
      \edef\AMSa{\hexnumber\amsa@group}
      \makeatother
      \mathchardef\upi="0\UPM19
      \mathchardef\umu="0\UPM16
      \mathchardef\upartial="0\UPM40
      \mathchardef\leqslant="3\AMSa36
      \mathchardef\geqslant="3\AMSa3E
    \fi
  \fi
\fi % End of NFSS release 1

\ifnfsstwo
  \DeclareMathAlphabet{\mathbfit}{OT1}{cmr}{bx}{it}
  \SetMathAlphabet\mathbfit{bold}{OT1}{cmr}{bx}{it}
  \DeclareMathAlphabet{\mathbfss}{OT1}{cmss}{bx}{n}
  \SetMathAlphabet\mathbfss{bold}{OT1}{cmss}{bx}{n}
  \ifAMStwofonts
    \ifCUPmtlplainloaded \else
      \DeclareSymbolFont{UPM}{U}{eur}{m}{n}
      \SetSymbolFont{UPM}{bold}{U}{eur}{b}{n}
      \DeclareSymbolFont{AMSa}{U}{msa}{m}{n}
      \DeclareMathSymbol{\upi}{0}{UPM}{"19}
      \DeclareMathSymbol{\umu}{0}{UPM}{"16}
      \DeclareMathSymbol{\upartial}{0}{UPM}{"40}
      \DeclareMathSymbol{\leqslant}{3}{AMSa}{"36}
      \DeclareMathSymbol{\geqslant}{3}{AMSa}{"3E}
    \fi
  \fi
\fi % End of NFSS release 2

\ifCUPmtlplainloaded \else
  \ifAMStwofonts \else % If no AMS fonts
    \def\upi{\pi}
    \def\umu{\mu}
    \def\upartial{\partial}
  \fi
\fi

\title
[Microwave background anisotropies due to non-linear structures]
{Microwave background anisotropies due to non-linear structures in
open and $\Lambda$ universes}

\author[Y.~Dabrowski et al.]
{Y.~Dabrowski\thanks{Email: youri@mrao.cam.ac.uk},
M.J.~Hall, I.L.~Sawicki and A.N.~Lasenby\\
Astrophysics Group,
Cavendish Laboratory,
Madingley Road,
Cambridge,
CB3 0HE,
UK}
\date{Accepted ???. Received ???; in original form \today}
\pagerange{\pageref{firstpage}--\pageref{lastpage}}
\pubyear{1997}

\begin{document}
\newcommand {\modela}{{\it this work (1)}}
\newcommand {\modelb}{{\it this work (2)}}
\newcommand {\modelc}{{\it this work (3)}}
\newcommand {\arccosh}{{\rm arccosh}}
\def\pdiff#1#2{{\frac{\partial #1}{\partial #2}}}
\def\Ddiff#1#2{{\frac{{\rm d} #1}{{\rm d} #2}}}
\newcommand {\OPEN}  {{\it Open} }
\newcommand {\FLATL} {{\it Flat-$\Lambda$} }
\newcommand {\FLAT}  {{\it EdS} }
\newcommand {\FLATC} {{\it EdS}}
\newcommand {\FLATLONG}{{\it Einstein de Sitter} }

\maketitle
\label{firstpage}

\begin{abstract}
A new method arising from a gauge-theoretic approach to general
relativity is applied to the formation of clusters in an expanding
universe.  The three cosmological models ($\Omega_0$=1,
$\Omega_{\Lambda}$=0), ($\Omega_0$=0.3, $\Omega_{\Lambda}$=0.7),
($\Omega_0$=0.3, $\Omega_{\Lambda}$=0) are considered, which extends
our previous application (Lasenby et al. 1999, Dabrowski et al. 1999).
A simple initial velocity and density perturbation of finite extent is
imposed at the epoch $z=1000$ and we investigate the subsequent
evolution of the density and velocity fields for clusters observed at
redshifts $z=1$, $z=2$ and $z=3$.  Photon geodesics and redshifts are
also calculated so that the Cosmic Microwave Background (CMB)
anisotropies due to collapsing clusters can be estimated.  We find
that the central CMB temperature decrement is slightly stronger and
extends to larger angular scales in the non-zero $\Omega_{\Lambda}$
case.  This effect is strongly enhanced in the open case.
Gravitational lensing effects are also considered and we apply our
model to the reported microwave decrement observed towards the quasar
pair PC~1643+4631~A\&B.

\end{abstract}


\section{Secondary Gravitational Effect}
\label{sec:secondary}
Rees and Sciama (1968) suggested that the presence of an evolving
structure on the line of sight of a Cosmic Microwave Background (CMB)
photon could significantly affect its observed temperature. This
secondary gravitational effect is often described in terms of the
potential well experienced by the CMB photon. For example, the
potential well of a collapsing cluster becomes deeper over time so
that the CMB photon has to {\it climb out} of a well deeper than that
into which it {\it fell}, suffering a net loss of energy.  As
suggested in Rees and Sciama (1968) there is however a competing
effect due to the extra time delay encountered by the photon.  The
overall effect can therefore be of either sign.  The photon
accumulates a redshift along its geodesic, resulting in a net
temperature perturbation. In the weak-field approximation we have (see
Mart\'{\i}nez-Gonz\'alez, Sanz \& Silk 1990)
\begin{equation}
\label{dphidt}
\frac{\Delta T}{T}=-\frac{2}{c^2}\int\frac{\partial \Phi}{\partial t}dt,
\end{equation}
where $\Phi$ is the gravitational potential of the perturbation and
$t$ is the cosmic time. A rough estimate of an upper limit to this
effect can be estimated by assuming that the potential well varies
from $\Phi=0$ to $\Phi=\Phi_c$ during the time that the photon
traverses it, where $\Phi_c$ is the gravitational potential of a rich
Abell cluster. In this case, we have
\begin{equation}
\frac{\Delta T}{T}<-\frac{2}{c^2}\Phi_c.
\end{equation}
The potential of the cluster can be related to the velocity dispersion
$\sigma$ through $\sigma^2=GM/R=\Phi_c$, where $M$ and $R$ are the
mass and radius of the cluster. Assuming $\sigma = 1000$~km~s$^{-1}$
gives
\begin{equation}
\frac{\Delta T}{T} < -2.2\times10^{-5}.
\end{equation}
In most cases, this anisotropy should therefore be small compared with
other secondary anisotropies caused by the interaction of CMB photons
with non-linear structures, such as the thermal Sunyaev-Zel'dovich
(SZ) effect.  However, equation~(\ref{dphidt}) is only valid in the
linear regime and a fully relativistic treatment in the non-linear
regime is necessary to estimate accurately the Rees-Sciama effect.

Several methods have been used to model the formation of galaxy
clusters in an expanding universe. The modelling of arbitrary
perturbations in a fully general relativistic manner has not proved
possible due to the complexity of the calculations involved.  The most
popular method has been the linearised approach (equation
\ref{dphidt}), allowing the modelling of arbitrarily shaped
inhomogeneities (Mart\'{\i}nez-Gonz\'alez \& Sanz 1990;
Mart\'{\i}nez-Gonz\'alez et al. 1990; Chodorowski 1991). This method
has been particularly used for Great Attractor-like structures located
at redshifts as high as $z=5.9$, where less strong non-linear effects
are expected (S\'aez, Arnau \& Fullana 1993, 1995; Arnau,
Fullana \& S\'aez 1994; Fullana, S\'aez \& Arnau 1994).

In this paper we are interesting in modelling rich galaxy clusters
which are often ellipsoid or even quasi spherical (e.g. Coma
cluster). This leads to spherical symmetry as a natural approximation,
which allows a general relativistic derivation.  The earliest attempts
of modelling are of the {\it Swiss Cheese} type, where the collapsing
Friedmann-Robertson-Walker (FRW) region is compensated for by a
surrounding region of vacuum matched onto the expanding FRW universe
(Rees \& Sciama 1968; Nottale 1982; Nottale 1984). In these models,
analytic calculations can be performed in the fully non-linear regime
and the observer can be located in the external, unperturbed
universe. However, the matter distribution is unrealistic.

More recent models have used the Tolman-Bondi solution for a
pressureless, spherically symmetric matter distribution to provide a
more realistic density perturbation (Panek 1992; Quilis, Ib\'a\~nez \&
S\'aez 1995; Quilis and S\'aez 1998).  These, however, are rarely
compensated and comoving observers are not in a region modelled by a
homogeneous FRW metric, which adds difficulty in estimating the
temperature anisotropy.

Finally, a new model arising from a gauge-theoretic approach to
gravity was proposed by Lasenby et al. (1999).  This method employs a
gauge, mainly characterised by a unique global time coordinate, which
reduces the equations to an essentially Newtonian form.  The model
avoids the problem of streamline crossing, while keeping the density
profile compensated and realistic. Furthermore, the perturbation is of
finite size, allowing the observer to be placed in a non-perturbed
region.

In this paper we extend this new theoretical approach to the case of a
non-zero cosmological constant $\Lambda$ and to the case of open
cosmologies.  The non-zero $\Lambda$ model is of particular interest
since recent observations of high redshift Type Ia supernovae are
inconsistent with $\Lambda=0$ flat or open cosmologies, suggesting
that our universe may be accelerating (Perlmutter et al. 1997, 1999).
When the Type Ia supernovae results are combined with observations of
the CMB, an approximately flat FRW cosmological model is suggested
with $\Omega_0\approx 0.3$ and $\Omega_{\Lambda} \approx 0.7$
(Efstathiou et al. 1999).  $\Omega_0$ is the mass density parameter at
the present epoch and $\Omega_{\Lambda}=\Lambda/3H_0^2$, where $H_0$
is the Hubble constant at the present time.

\section{Assumptions}
\label{sec:assumptions}

Three cosmological models are considered here: The \FLATLONG (\FLATC),
\FLATL and \OPEN models, corresponding to ($\Omega_0$=1, $\Omega_{\Lambda}$=0),
($\Omega_0$=0.3, $\Omega_{\Lambda}$=0.7), and ($\Omega_0$=0.3,
$\Omega_{\Lambda}$=0) respectively.
%Unless stated otherwise, we
%perform simulations for $H(t_0)=100$~km~s$^{-1}$Mpc$^{-1}$, where
%$t_0$ is the present time.
Results will be quoted using the reduced Hubble parameter $h(t)$,
which is equal to $H(t)$ expressed in units of
$100$~km~s$^{-1}$Mpc$^{-1}$.  Furthermore, quantities evaluated at the
present time $t=t_0$ will be written simply with a zero subscript,
e.g., $h(t_0)=h_0$. Throughout this work we take $t_i$, the time at
which the initial perturbation is applied, to represent the epoch
$z=1000$. We employ natural units $G=c=\hbar=1$, unless stated
otherwise.

Throughout this paper, we assume that the cosmological fluid as well
as the evolving non-linear structure are pressureless. Quilis et al.
(1995) included a hot gas component in their model and found that the
pressure effects are in fact negligible and that the collapse is well
approximated by the pressureless assumption.

As mentioned above, the perturbation is assumed to be spherically
symmetric which, combined with the pressureless assumption, leads to
the unrealistic situation where the structure collapses to form a
singularity. As a result, the estimates of CMB anisotropies given in
this work should be regarded as upper limits. However, the initial
perturbations are chosen so that the CMB photon traverses the cluster
before it becomes singular. Throughout this paper, the maximum
baryonic density experienced by the photon at the centre of the
non-linear structure is set to $10^4h_0^{1/2}~{\rm protons~m^{-3}}$.
Furthermore the clusters considered in this work are located at high
redshifts ($z=1$, $z=2$ and $z=3$) and should therefore still be in
the process of formation. In these cases the large infalling motions
predicted by the spherically symmetric pressureless assumptions may be
less unrealistic.

We further assume that, at all times, the baryon component contributes
$10h_0^{-3/2}$ per cent of the total mass, the remainder being dark
matter. The $h$-dependence chosen here ensures that the total
gravitational mass scales as $h_0^{-1}$, as expected within the
isothermal sphere approximation (e.g. Peebles 1993). In this case the
total density follows the power law form $\rho\propto r^{-2}$ and the
baryon mass alone scales as $h_0^{-5/2}$.

\section{Fluid Dynamics}
This paper follows the theoretical approach of Lasenby et al. (1999)
and Lasenby, Doran \& Gull (1998) where exact and fully relativistic
derivations are given using a new gauge-theoretic approach to gravity.
The resulting field equations are remarkably straightforward and are
here presented as if derived from classical first principles (see
Lasenby et al. (1997) for rigorous calculations).

The forming cluster and the external universe are both described by
the same dynamical equations, the centre of the cluster being at the
origin of the spherical coordinates ($r$, $\theta$, $\phi$). Since we
are only concerned with spherically symmetric systems, we choose
hereafter $\theta = \pi/2$ without loss of generality.  The symbol $t$
is used for the dynamical time.

The two scalar quantities needed to describe the fluid dynamics are
the density $\rho (t,r)$ and the radial velocity
\begin{equation}
\label{eq:u_definition}
u(t,r) = \frac{{\rm d}r}{{\rm d}t}
\end{equation}
along matter geodesics. The total gravitational mass $M(t,r)$
contained within a sphere of radius r is defined by
\begin{equation}
\label{eq:dM_dr}
\left(\frac{\partial M}{\partial r}\right)_t = 4\pi r^2\rho,
\end{equation}
or
\begin{equation}
\label{eq:mass_definition}
M(t,r) = \int^r_0 4\pi r'^2\rho(t,r')dr'.
\end{equation}
The mass of fluid flowing through a sphere of radius $r$ per unit time
is given by
\begin{equation}
\label{eq:dM_dt}
\left(\frac{\partial M}{\partial t}\right)_r = -4\pi r^2\rho u.
\end{equation}
One can define the streamline derivative (i.e. co-moving with the
fluid) as
\begin{equation}
\label{eq:d_dl}
\frac{{\rm d}}{{\rm d}l} \equiv \frac{\partial}{\partial t} +
u\frac{\partial}{\partial r}.
\end{equation}
We can see from (\ref{eq:dM_dr}) and (\ref{eq:dM_dt}) that the total
gravitational mass enclosed is conserved along the fluid streamlines,
\begin{equation}
\label{eq:dM_dl}
\frac{{\rm d}M}{{\rm d}l} = 0,
\end{equation}
which forbids the possibility of streamline crossing.

\subsection{Equation of Continuity}
By equating the total mass of fluid flowing out of a given volume per
unit time to the decrease per unit time of fluid in the same volume,
one obtains the standard {\it equation of continuity}
\begin{equation}
\label{eq:continuity}
\frac{{\rm d}\rho}{{\rm d}l} =
\frac{\partial \rho}{\partial t} +
u \frac{\partial \rho}{\partial r} = 
-\left(\frac{2u}{r}+H\right)\rho,
\end{equation}
where 
\begin{equation}
\label{eq:H_definition}
H(t,r) \equiv \frac{\partial u(t,r)}{\partial r}.
\end{equation}

\subsection{Euler's Equation}
Newton's second law applied on a fluid particle gives the usual
{\it Euler's equation}
\begin{equation}
\label{eq:Euler0}
\frac{\partial u}{\partial t} + u \frac{\partial u}{\partial r} =
F(t,r),
\end{equation}
where $F$ is the total force applied on the particle. $F$ includes the
radial gravitational force $F_g = -M/r^2$ as well as the repulsive
force $F_{\Lambda}$ proportional to $r$ due to the cosmological
constant $\Lambda$. To agree with the usual convention we set the
constant of proportionality to $\Lambda/3$ so that
$F_{\Lambda} = (\Lambda/3)r$.  Finally equation~(\ref{eq:Euler0}) reads
\begin{equation}
\label{eq:Euler}
\frac{{\rm d}u}{{\rm d}l} =
\frac{\partial u}{\partial t} + u \frac{\partial u}{\partial r} =
-\frac{M}{r^2}+\frac{\Lambda}{3}r.
\end{equation}

\subsection{Bernoulli's Equation}
Multiplying Euler's equation (\ref{eq:Euler}) by $u$ and integrating
along the streamline ${\rm d}l$, gives
\begin{equation}
\label{eq:Bernoulli}
\frac{1}{2}u^2 - \left(\frac{M}{r} + \frac{\Lambda}{6}r^2\right) = E(t,r),
\end{equation}
where the constant of integration $E(t,r)$ is constant along the fluid
streamlines, i.e.
\begin{equation}
\label{eq:dE_ds}
\frac{{\rm d}E}{{\rm d}l} = 0.
\end{equation}
In equation (\ref{eq:Bernoulli}), $u^2/2$ can be regarded as the
kinetic energy of a fluid particle, while $-M/r - (\Lambda/6)r^2$ can
be thought of as its potential energy composed of the usual
gravitational potential and a potential proportional to $r^2$ due to
the presence of a cosmological constant.  $E(t,r)$ can therefore be
associated with the total energy of a fluid shell. If $E < 0$ the
shell will eventually collapse at the origin while if $E > 0$ the
shell will remain in expansion.  In other words, equation
(\ref{eq:Bernoulli}) can be regarded as a generalised {\it Friedman
equation} for which each fluid shell is allowed a different constant
$E$.

\subsection{Newtonian Gauge}
\label{sec:newtonian}
The physical significance of the above set of equations is remarkably
straightforward yet, as shown in Lasenby et al. (1998), they are fully
consistent with general relativity. This is because the degrees of
freedom offered by the gauge theoretic approach have been carefully
chosen so that the resulting equations look Newtonian. In particular
the gauge fields have been fixed so that the coordinate $r$ is a
measurable distance scale related to the strength of the tidal force,
and the coordinate $t$ is a measure of the proper time $\tau$ for all
observers comoving with the fluid, therefore ${\rm d}t$ = ${\rm d}l$ =
${\rm d}\tau$, $l$ being the affine parameter of the streamlines
(equation \ref{eq:d_dl}). Therefore this gauge choice enables us to
recover a global ``Newtonian'' time on which all observers can
agree. Lasenby et al. (1998) have named this gauge the {\it Newtonian
Gauge}. The associated line element is given by
\begin{eqnarray}
\label{eq:ds2}
{\rm d}s^2 &=& \left(1-\frac{u^2}{2E+1}\right){\rm d}r^2 +
\frac{2u}{2E+1}{\rm d}t{\rm d}r -
\frac{1}{2E+1}{\rm d}r^2 \nonumber \\
& & - r^2\left({\rm d}\theta^2 + \sin^2\theta{\rm d}\phi^2\right).
\end{eqnarray}
Similar analytic solutions have been found by Tolman (1934) and Bondi
(1947), however their use of $M(r)$ as the radial coordinate rather
than $r$ has the disadvantage of hiding the physical meaning of the
field equation and complicates the choice of initial conditions.

In homogeneous regions the fluid behaves as for standard cosmologies
and the density $\rho$ is a function of time only, so we have
\begin{equation}
\bar{M}(t,r)=\frac{4\pi}{3}r^3\bar{\rho}(t),
\end{equation}
where the overbar denotes quantities in the external homogeneous
region.  Here, equations (\ref{eq:dM_dl}), (\ref{eq:continuity}) and
(\ref{eq:Euler}) lead to
\begin{equation}
\frac{{\rm d}\bar{\rho}}{{\rm d}t} = -3\bar{H}\bar{\rho}(t),
\end{equation}
\begin{equation}
\bar{H}(t) = \frac{\bar{u}}{r},
\end{equation}
and
\begin{equation}
\frac{{\rm d}\bar{H}}{{\rm d}t}+\bar{H}^2 = -\frac{4\pi}{3}\bar{\rho} + \frac{\Lambda}{3}.
\end{equation}
It is clear from those standard results that we can identify $\bar{H}$
as the Hubble constant, and that the time parameter $t$ of the
Newtonian gauge agrees with the cosmic time in homogeneous regions of
our model. We also have
\begin{equation}
\label{eq:8pirho}
\frac{8\pi}{3}\bar{\rho}(t) = \bar{H}(t)^2\Omega(t),
\end{equation}
for the \FLAT and \OPEN models and
\begin{equation}
\label{eq:8pirho_l}
\frac{8\pi}{3}\bar{\rho}(t) = \bar{H}^2(t) - \frac{1}{3}\Lambda,
\end{equation}
for the \FLATL model.

\subsection{Streamline Equations}

In general, the solution to equation (\ref{eq:Bernoulli}) can be
expressed in terms of elliptic integrals. We choose here to integrate
numerically this equation using the following system of two first
order ordinary differential equations composed of
(\ref{eq:u_definition}) and Euler's equation:
\begin{eqnarray}
\label{eq:friedmansys}
\frac{{\rm d}r}{{\rm d}t} &=& u \nonumber \\
\frac{{\rm d}u}{{\rm d}t} &=& -\frac{M(r_i)}{r^2} + \frac{\Lambda}{3}r,
\end{eqnarray}
where $r_i$ is the fluid particle radius on a given streamline, at an
initial time $t_i$. This system allows us to compute the position $r$
of the fluid particle as well as its velocity $u$ at a later time $t$.
We note that, in this manner, the change from positive $u$ to negative
$u$ is automatically evaluated in the case of closed streamlines.  The
fluid density $\rho(t,r)$ and velocity gradient $H(t,r)$ are
calculated by differentiating numerically equations (\ref{eq:dM_dr})
and (\ref{eq:H_definition}).

However, equation (\ref{eq:Bernoulli}) has analytical solutions in two
particular cases. (i) For $\Lambda = 0$, which corresponds to the
standard Friedman models. This case is treated in detail in Lasenby et
al. (1999). (ii) For $E = 0$, where the solution is given by the
following streamline equation
\begin{eqnarray}
\label{eq:solution_flat}
t = t_i &+& \frac{1}{\sqrt{3\Lambda}}\arccosh\left(\frac{\Lambda}{3M(r_i)}r^3
+1\right) \nonumber \\
&-&\frac{1}{\sqrt{3\Lambda}}\arccosh\left(\frac{\Lambda}{3M(r_i)}r_i^3
+1\right).
\end{eqnarray}
In this case, the value of the velocity along the streamline is given by
(\ref{eq:Bernoulli}) itself:
\begin{equation}
\label{eq:vel}
u = \sqrt{ \frac{2M(r_i)}{r} + \frac{\Lambda}{3}r^2 }.
\end{equation}
We note that here $u$ is always positive since $E=0$.
Following Lasenby et al. (1999), the density is given by
\begin{equation}
\label{eq:density}
\rho(t,r) = \left(\frac{\partial r_i}{\partial r}\right)_t\frac{r_i^2}{r^2}
\rho(r_i),
\end{equation}
where
\begin{eqnarray}
\label{eq:dridr}
\left(\frac{\partial r_i}{\partial r}\right)_t^{-1} =
\frac{1}{Mr^{\frac{1}{2}}\sqrt{\Lambda r_i^3+6M}}
\Big[
Mr_i^{\frac{1}{2}}\sqrt{\Lambda r^3+6M} + \nonumber \\
\frac{4\pi}{3}r_i^2\rho(r_i)
\left(
r^{\frac{3}{2}}\sqrt{\Lambda r_i^3+6M} -
r_i^{\frac{3}{2}}\sqrt{\Lambda r^3+6M}
\right)
\Big].
\end{eqnarray}
The velocity gradient is found by differentiating
equation~(\ref{eq:Bernoulli}), which leads to
\begin{equation}
\label{eq:dudr}
\frac{\partial u}{\partial r} = 
-\frac{M}{r^2u}+\Lambda\frac{r}{3u}+\frac{1}{u}\left(
\frac{\partial r_i}{\partial r}\right)_t
\left[
\frac{1}{r}\frac{{\rm d}M}{{\rm d}r_i}+
\frac{{\rm d}E}{{\rm d}r_i}
\right].
\end{equation}

\section{Cluster Formation Model}
The fluid equations given above control the evolution of the external
homogeneous cosmology as well as the formation of the central
cluster. The initial perturbation is applied at an early time $t_i$
that we choose here to represent the epoch $z=10^3$. The perturbation
is of finite extent and we therefore need to fix initial conditions so
that standard cosmology is satisfied outside a given radius $R_i$.
Here we will use a family of simple four-parameter models based on
polynomial perturbations in the density and velocity fields, as
described in detail in Lasenby et al. (1999) and Dabrowski et
al. (1999).  The parameters are (i) the width of the perturbation
$R_i$; (ii) the velocity gradient at the origin $H(t_i,0)$; (iii) the
degree of the polynomial describing the perturbation, so that at
$r=R_i$, the velocity and its first $m$ derivatives match the
external values; (iv) the external velocity gradient $\bar{H}(t_i)$
equal to the Hubble constant at the time $t_i$. As a typical value, we
assume in this paper $m=3$.

\subsection{Linearised Equations}
The perturbation responsible for the formation of clusters of galaxies
is assumed to have grown from primordial fluctuations in the very
early universe. At the epoch $z=10^3$, the fluid dynamics should still
be well described by considering evolution in the linear regime.

We introduce the linearised variables
\begin{eqnarray}
	\label{eq:deltau}
	\delta u &\equiv& u - r \bar{H}(t)\nonumber \\
	\delta \rho & \equiv& \rho - \bar{\rho}(t) \\
	\delta M & \equiv& \int^r_0 4\pi s^2 \delta \rho(t,s)ds \nonumber \,,
\end{eqnarray}
where the overbarred quantities denote background homogeneous universe values.
Differentiating equation~(\ref{eq:deltau}) gives
\begin{equation} \label{eq:Ddeltau}
	\Ddiff{\delta u}{t} = -\frac{\delta M}{r^2} - \bar{H}\delta u\,,
\end{equation}
as in the $\Lambda = 0$ case (see Lasenby et al. 1999).
Differentiating again and ignoring second order terms yields
\begin{equation}
	\Ddiff{}{t}\left(\Ddiff{\delta u}{t}\right) + 
		3 \bar{H}\Ddiff{\delta u}{t}
		- \bar{H}^2 \delta u + \Lambda \delta u = 0 \,,
\end{equation}
where the background universe is assumed to be spatially flat ($E =
0$).  By parameterising $\delta u$ in terms of streamline starting
point, $r_i$ and cosmic time, $t$, the streamline derivatives become
derivatives with respect to $t$:
\begin{equation}
	\label{eq:ddu}
	\delta \ddot{u} + 3 \bar{H} \delta \dot{u} - (\bar{H}^2 - \Lambda)
	\delta u = 0\,,
\end{equation}
where the overdot denotes derivatives with respect to $t$.  In the
limit of small $t$ and $\Lambda$
\begin{equation}
	\bar{H} \approx \frac{2}{3t} + \frac{\Lambda}{6} t,
\end{equation}
and $\delta u$ goes as
\begin{equation}
	\delta u = A{t^{-4/3}} + Bt^{1/3} + O(t^{2/3})\,.
\end{equation}
Assuming that we are in an epoch where the decaying mode $t^{-4/3}$
can be ignored but sufficiently early that the $t^{2/3}$ term is not
yet significant, then
\begin{equation}
	\Ddiff{\delta u}{t}=\frac{1}{2}\bar{H}\delta u\,.
\end{equation}
Alternatively, it can be said that at early enough times,
$\bar{H}^2\gg\Lambda$ and so the $\Lambda$ term may be ignored in
equation~(\ref{eq:ddu}). Thus using equation~(\ref{eq:Ddeltau}) we
have
\begin{equation} \label{eq:initconst}
	\frac{3}{2}\bar{H}\delta u = - \frac{\delta M}{r^2}\,,
\end{equation}
or expressed in terms of the perturbed and un-perturbed quantities
\begin{equation} \label{eq:initconst2}
u = r\bar{H} -\frac{2M}{3\bar{H}r^2}+\frac{r}{3\bar{H}}\frac{8\pi}{3}
\bar{\rho}.
\end{equation}
Substituting $8\pi\bar{\rho}/3$ with values of equations
(\ref{eq:8pirho}) and (\ref{eq:8pirho_l}) gives an extra condition
on the initial data
\begin{eqnarray}
\label{eq:init_vel}
u(t_i,r) = \frac{2r}{3\bar{H}}\left(\frac{3+\Omega_0}{2}\bar{H}^2-\frac{M}{r^3}\right),
\end{eqnarray}
for the \FLAT and \OPEN models and
\begin{eqnarray}
\label{eq:init_vel_l}
u(t_i,r) = \frac{2r}{3\bar{H}}\left(2\bar{H}^2-\frac{M}{r^3}
	-\frac{1}{6}\Lambda\right),
\end{eqnarray}
for the \FLATL model.  The initial density perturbation can now be
derived from the initial velocity field as follows: Differentiating
(\ref{eq:init_vel}) and (\ref{eq:init_vel_l}) with respect to $r$
gives
\begin{equation}
\label{eq:init_density}
\rho(t_i, r) = \frac{3\bar{H}}{8\pi}\left[
	\left(3+\Omega_0\right)\bar{H}-\frac{2u}{r}-\left(\frac{\partial u}{\partial r}\right)_{t_i}
\right],
\end{equation}
for the \FLAT and \OPEN cosmologies, and
\begin{equation}
\label{eq:init_density_l}
\rho(t_i, r) = \frac{3\bar{H}}{8\pi}\left[
	4\bar{H}-\frac{2u}{r}-\left(\frac{\partial u}{\partial r}\right)_{t_i}-\frac{1}{3\bar{H}}\Lambda
\right],
\end{equation}
for the \FLATL case.

\subsection{Initial Conditions}
\label{sec:initial_conditions}
\begin{figure}
\centerline{\epsfig{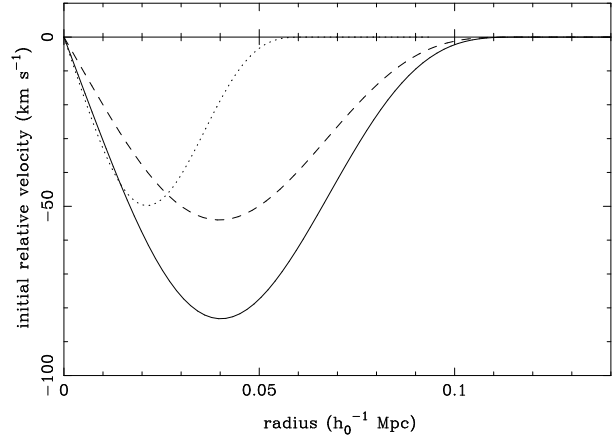}}
\caption{Initial relative velocity $u(t_i,r) - \bar{H}(ti)r$ where $t_i$
represents the epoch $z=10^3$. The solid, dashed and dotted lines are
for the \OPEN, \FLATL ($\Omega_{\Lambda}=0.7$) and \FLAT models
respectively.}
\label{fig:init_velocity}
\end{figure}

\begin{figure}
\centerline{\epsfig{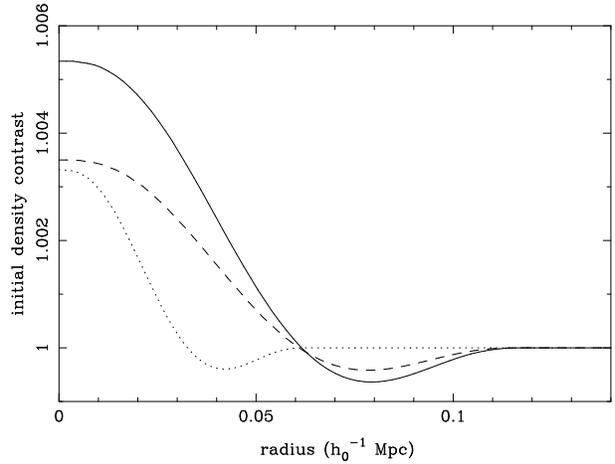}}
\caption{As in Fig.~\ref{fig:init_velocity}, but for the initial
density contrast $\rho(t_i,r)/\bar{\rho}(t_i)$.}
\label{fig:init_density}
\end{figure}

At the time $t_i$, the initial conditions are defined completely by
the 4 parameter velocity profile while the corresponding initial
density profile is given by equation~(\ref{eq:init_density}) or
(\ref{eq:init_density_l}).  The external universe $\bar{H}(t_i)$ is
fully defined by the desired cosmology (i.e. $\Omega_0$,
$\Omega_{\Lambda}$ and $h_0$). We have for the \FLAT and \OPEN models
\begin{equation}
\bar{H}(t_i) = H_0( 1 + z_i ) \sqrt{ 1 + \Omega_0  z_i },
\end{equation}
and for the \FLATL cosmology
\begin{equation}
\bar{H}(t_i) = \sqrt{\left(1+z_i\right)^3 \left(H_0^2 -
	\frac{1}{3}\Lambda\right) + \frac{1}{3}\Lambda}.
\end{equation}
The two remaining parameters $R_i$ and $H(t_i,0)$ are fixed such that
the resulting cluster is similar to those observed today, using the
following characteristics: the redshift of the cluster $z_c$, its core
radius $R_c$ (where $R_c$ is defined as the radius at which the
cluster density falls to one-half its maximum value) and its central
baryonic density $\rho_c$. Because we are interested in modelling rich
clusters, the core radius is fixed to $R_c=0.23h_0^{-1}$~Mpc and the
central density is taken to be $\rho_c=10^4h_0^{1/2}$~${\rm
p~m}^{-3}$.  The $h$-dependences of $R_c$ and $\rho_c$ are chosen so
that the cluster's observed angular size and X-ray flux are
independent of $h_0$ (Jones \& Forman 1984; Peebles
1993). Figs.~\ref{fig:init_velocity} and
\ref{fig:init_density} show the initial velocity and density profiles
in the \OPEN, \FLATL and \FLAT models. In all three cases, the
resulting cluster observed at $z_c=1$ satisfies the same properties
$R_c= 0.23h_0^{-1}$~Mpc and $\rho_c=10^4h_0^{1/2}$~${\rm p~m}^{-3}$.
We note that low-$\Omega_0$ models require larger and deeper
perturbation. This can be explained as follows: The equation
equivalent to (\ref{eq:ddu}) but for the density perturbation
evolution is given by
\begin{equation}
\label{eq:ddrho}
\ddot{\delta}=4\pi\bar{\rho}\delta - 2\bar{H}\dot{\delta},
\end{equation}
where $\delta=\delta\rho/\bar{\rho}$. The first term of the right hand
side governs the growth of the perturbation while the second can be
regarded as a damping term due to the drag from the expansion of the
universe. In the low density universe cases (\FLATL and \OPEN models),
the growth term is smaller than in the critical density \FLAT model.
Furthermore, $\bar{H}$ is larger in the \FLATL and \OPEN cosmologies
(see equations \ref{eq:8pirho} and \ref{eq:8pirho_l}), enhancing the
damping term. Therefore it is to be expected that the initial
perturbations in the case of the \FLATL and \OPEN models have to be
larger than in the \FLAT case in order to give rise to similar
structures at a later time. A more detailed study shows that in the
\FLATL model the growth suppression is less marked than for the \OPEN
model (see Peacock 1999). This explains why the initial perturbation
of the \FLATL case is less pronounced than for the \OPEN case, as seen
in Figs.~\ref{fig:init_velocity} and \ref{fig:init_density}.

\subsection{Cluster Characteristics}
As discussed in Section~\ref{sec:assumptions}, the pressureless and
spherically symmetric assumptions inevitably lead to non linear
structures with large infall velocities, which would probably be
realistic only for distant clusters still in the process of
formation. Clusters observed at redshift $z=1$, $z=2$ and $z=3$ will
be considered here and in the following sections.

\begin{figure}
\centerline{\epsfig{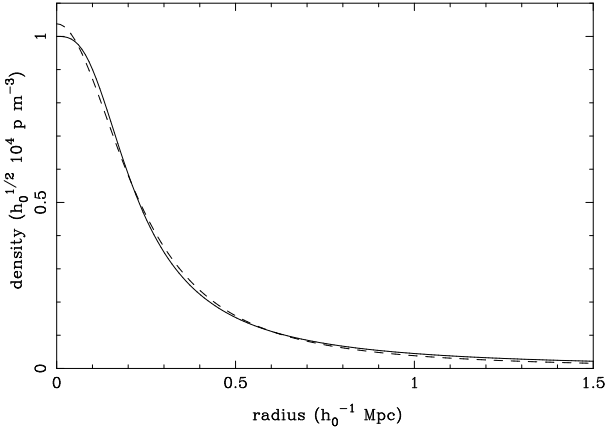}}
\caption{The baryonic density of the cluster as a function of radius
at the time corresponding to the redshift $z=1$. The dashed line is
the best-fitting $\beta$-model up to $r=1.5 h_0^{-1}$~Mpc. The \OPEN
model has been assumed here (see text).}
\label{fig:central_density}
\end{figure}

The central density profile obtained at $z=1$ when applying the
initial conditions of Section~\ref{sec:initial_conditions} is given in
Fig.~\ref{fig:central_density} up to the $1.5 h_0^{-1}$~Mpc Abell
radius, where the \OPEN model is assumed.
We found that the central density profiles obtained in the context of
the \FLAT and \FLATL cosmologies are almost indistinguishable from
that of the \OPEN case.  Furthermore, varying the redshift at which
the cluster is observed from $z=1$ to $z=3$ does not affect the
central density profile either.  It is remarkable that by only fixing
the desired central density $\rho_c$ and core radius $R_c$, the
resulting density profile is almost identical in any cosmologies and
at any redshift despite the significant difference between the initial
perturbations, as seen in Figs.~\ref{fig:init_velocity} and
\ref{fig:init_density}.  However, it was found in Dabrowski et
al. (1999) that the shape of the density profile is sensitive to the
other free parameter of our model $m$, which is here set to $m=3$.

As discussed in detail in Dabrowski et al. (1999), the obtained
density profile is well fitted by $\beta$-models
\begin{equation}
\rho(r)=\frac{\rho_0}
{\left[1+\left(\frac{r}{R}\right)^2\right]^{\frac{3}{2}\beta}},
\end{equation}
usually used to fit observed profiles and rotation curves.  It was
also found that density distributions such as those derived from
N-body codes (Navarro Frenk \& White 1996, 1997, hereafter NFW) give
good agreement as regard the corresponding mass distribution of the
cluster.  However, the best least-square fit is obtained for a
$\beta$-model with $\beta=2/3$. Indeed, as mentioned in
Section~{\ref{sec:assumptions} our model predicts density profiles
$\rho \propto r^{-2}$ away from the origin which is more consistent
with a $\beta=2/3$ model than with the NFW density profiles which
behave like $r^{-3}$.  Another difference between the density profiles
obtained here and those of NFW is regarding their behaviour at the
centre of the cluster.  In this work the central density keeps a
finite value while it diverges to infinity in the case of NFW density
profiles.

%%
%\begin{figure}
%\centerline{\epsfig{
%file=fig/density.eps, angle=-90, width=8cm}}
%\caption{}
%\label{fig:density}
%\end{figure}
%

%\subsubsection{Mass Profile}
%
%\begin{figure}
%\centerline{\epsfig{
%file=fig/mass.eps, angle=-90, width=8cm}}
%\caption{}
%\label{fig:mass}
%\end{figure}
%

The velocity distribution, as evolved at the time where the cluster is
observed, is shown in Fig.~\ref{fig:velocity} for $z=1$ and for the
three \FLAT, \FLATL and \OPEN models.
\begin{figure}
\centerline{\epsfig{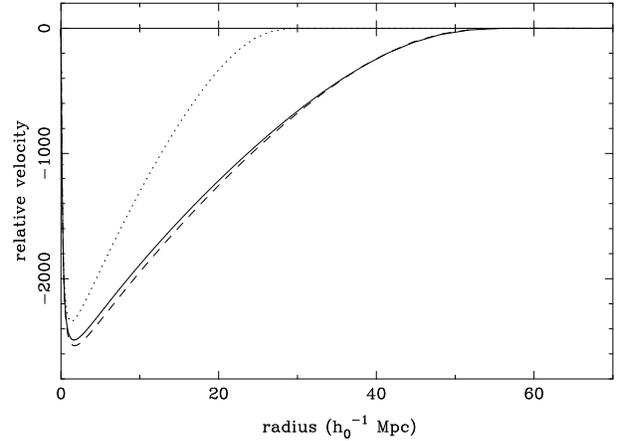}}
\caption{The velocity distribution relative to the Hubble flow
$u(t,r)-\bar{H}(t)r$ as a function of radius at the time corresponding
to the redshift $z=1$. The solid, dashed and dotted lines are for
the \OPEN, \FLATL and \FLAT cosmologies respectively.}
\label{fig:velocity}
\end{figure}
As expected from an initial perturbation of finite extent, the
velocity profile matches exactly that of the external homogeneous
universe after a finite radius. This effect can be seen clearly on
Fig.~\ref{fig:velocity} where the velocity relative to the Hubble flow
vanishes for $r>28 h_0^{-1}$~Mpc in the \FLAT cosmology and for $r>55
h_0^{-1}$~Mpc in the \FLATL and \OPEN cases. It is not surprising that
this radius is roughly a factor of 2 smaller in the \FLAT model since
the initial velocity and density perturbations were smaller in the
same proportion, as seen in Figs.~\ref{fig:init_velocity} and
~\ref{fig:init_density}.

Other quantities of particular interest are the turnaround radius
$R_t$, which is defined as the radius at which the fluid velocity
changes from being radially inwards to radially outwards, and the
ratio $\zeta$ of the cluster central density to that of the external
universe.  These characteristics, together with total gravitational
masses contained within spheres of given radii, are summarised in
Table~\ref{tab:characteristics} for clusters at redshifts $z=1$, $z=2$
and $z=3$ in the three \FLAT, \FLATL and \OPEN cosmologies. In all
cases, the observed clusters have the same central density $\rho_c =
10^4 h_0^{1/2}{\rm~p~m^{-3}}$ and core radius $R_c = 0.23
h_0^{-1}$~Mpc.
\begin{table}
\caption{Cluster properties. $z$ denotes the redshift at which the cluster
is observed (see text).  $R_t$ is the turnaround radius quoted in
$h_0^{-1}$~Mpc and $\zeta$ is the ratio of the central cluster density
to that of the external Universe.  $M_1$, $M_{1.5}$, $M_2$ and $M_4$
are total gravitational masses contained within spheres of radii
$r=1$, $1.5$, $2$ and $4 h_0^{-1}$~Mpc respectively. All masses are
quoted in $10^{15}h_0^{-1}~{\rm M_{\odot}}$.}
\begin{center}
\begin{tabular}{lccccccc} \hline
Model  & $z$ & $R_t$ & $\zeta$ & $M_1$ & $M_{1.5}$ & $M_2$ & $M_4$\\ \hline
\FLAT  &  1  & 6.26 & 1122 & 1.05 & 1.80 &  2.65  & 6.70\\
\FLATL &  1  & 10.74 & 3704 & 1.00 & 1.75 & 2.50 & 6.05\\
\OPEN  &  1  & 8.70 & 3704 & 1.00 & 1.70 & 2.50 & 6.05\\
\hline
\FLAT  &  2  & 3.22 & 332.5 & 1.05 & 1.90 & 2.90 & 7.90\\
\FLATL &  2  & 6.01 & 1097 & 1.00 & 1.85 & 2.65 & 6.75\\
\OPEN  &  2  & 4.94 & 1097 & 1.00 & 1.85 & 2.65 & 6.75\\
\hline
\FLAT  &  3  & 1.97 & 140.3 & 1.10 & 2.10 & 3.15 & 9.50\\
\FLATL &  3  & 3.82 & 463.0 & 1.05 & 1.90 & 2.80 & 7.50\\
\OPEN  &  3  & 3.23 & 463.0 & 1.05 & 1.90 & 2.80 & 7.55\\
\hline
\end{tabular}
\end{center}
\label{tab:characteristics}
\end{table}

\section{Photon Paths and Redshifts}
In order to estimate the effect of non linear structure formation on
the CMB we need to calculate the photon geodesics and redshifts in the
context of the metric given in equation (\ref{eq:ds2}), as carried out
in Lasenby et al. (1999). The photon motion is parametrised by the
Newtonian time parameter $t$ discussed in
Section~\ref{sec:newtonian}. The path is defined by four quantities,
the photon radius $r(t)$, position angle $\phi(t)$, 4-momentum
direction $\chi(t)$ and frequency $\omega(t)$.
\begin{figure}
\centerline{\epsfig{
file=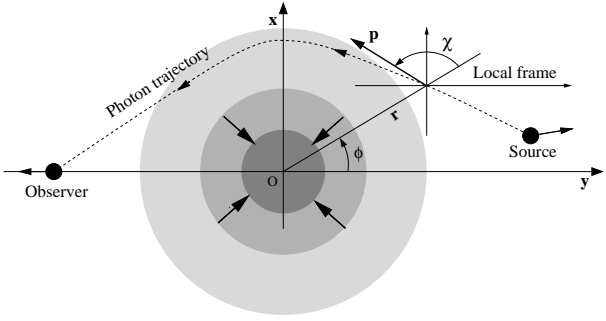,width=8cm}}
\caption{
Geometrical arrangement defining the photon geodesic integration.  The
grey discs denote the spherically symmetric collapsing cluster which
is located at the origin of the coordinate system. The photon is
emitted from a source co-moving with the Hubble flow. The photon path
is defined by the photon ($r$, $\phi$) position as well as the
direction $\chi$ and frequency $\omega$ of its 4-momentum $p$.}
\label{fig:geo}
\end{figure}
As shown in Fig.~\ref{fig:geo}, our model allows us to start a photon
path at a given source position co-moving with the Hubble flow;
integrate the geodesic equations through the homogeneous universe and
through the collapsing cluster; and finally collect the photon at the
observer position, co-moving with the Hubble flow as well.  The
calculation is identical in any spherically symmetric and pressure
free cosmological model so that the set of four first order
differential equation given in Lasenby et al. (1999) is still valid
here.

\section{Effect on the CMB}
As discussed in Section~{\ref{sec:secondary}, the presence of a time
dependent potential on the line of sight will affect our observation
of the CMB temperature. For the model considered here, Lasenby et al.
(1999) found
\begin{equation}
\label{eq:dt_t}
\frac{\Delta T}{T} = \int_{t_1}^{t_2}\left(-\pdiff{\Delta}{r}\cos^2\chi-\frac{\Delta}{r}\sin^2\chi\right){\rm d}t,
\end{equation}
where the integral is evaluated along the photon path.  $\Delta$ is
the fluid velocity relative to the Hubble flow
\begin{equation}
\Delta(t,r) = u(t,r) - r\bar{H}(t),
\end{equation}
$t_1$ is time at the surface of last scattering and $t_2$ is the time
at the observer. Numerical results are given in Table~\ref{tab:dt_t}
for CMB photons that travel straight through the centre of collapsing
clusters at $z=1$, $z=2$ and $z=3$.
\begin{table}
\caption{Estimates of the central temperature decrements $\Delta T/T$ in 
units of $10^{-5}$ for clusters at redshift $z=1$, $z=2$ and $z=3$.
All clusters have the same central density $\rho_c = 10^4
h_0^{1/2}{\rm~p~m^{-3}}$ and core radius $R_c = 0.23
h_0^{-1}$~Mpc. Three cosmological models are considered.  }
\begin{center}
\begin{tabular}{lccc} \hline
Model  & $\Delta T/T$, $z=1$ & $\Delta T/T$, $z=2$ & $\Delta T/T$, $z=3$\\ \hline
\FLAT  &  -0.64  & -0.45 & -0.32 \\
\FLATL &  -0.80  & -0.64 & -0.51 \\
\OPEN  &  -0.52  & -0.43 & -0.36 \\
\hline
\end{tabular}
\end{center}
\label{tab:dt_t}
\end{table}
The \FLAT, and \OPEN cosmological models considered here give roughly
the same predictions for the central temperature decrement.  We find
however that the central anisotropy is slightly stronger in the \FLATL
case.  When looking at the redshift dependence, Table.~\ref{tab:dt_t}
shows that, for a given cluster central density and core radius, the
further away the cluster is, the smaller the magnitude of the
Rees-Sciama effect is. At earlier epochs, collapsing structures are
less decoupled from the Hubble expansion and are therefore exposed to
less strong non-linear effects, therefore it is not surprising that
the effect gets dimmer.
\begin{figure*}
\centerline{
\epsfig{file=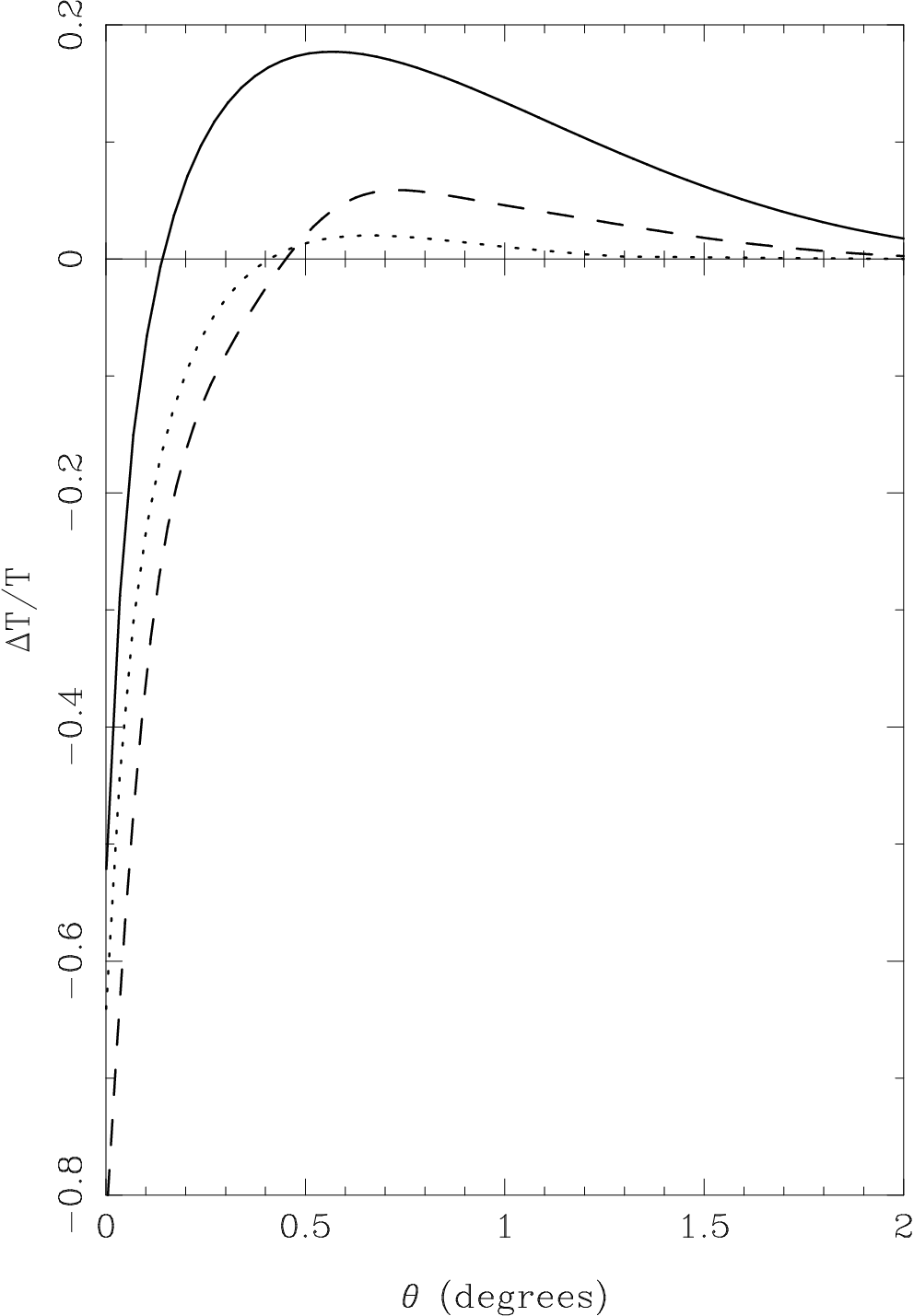, width=5.5cm, height=8.2cm}
\hspace{0.4cm} 
\epsfig{file=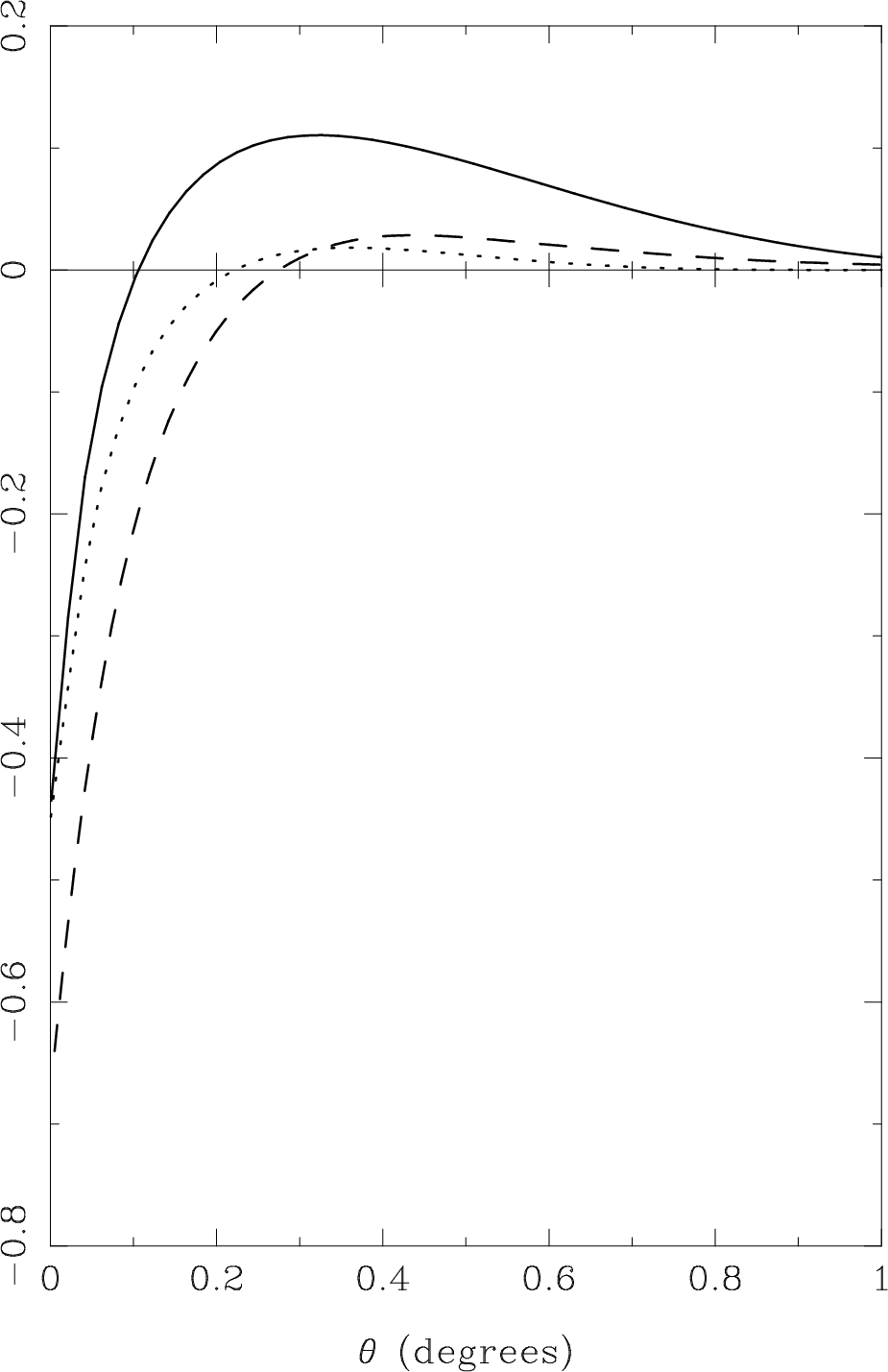, width=5.1cm, height=8.2cm}
\hspace{0.4cm}
\epsfig{file=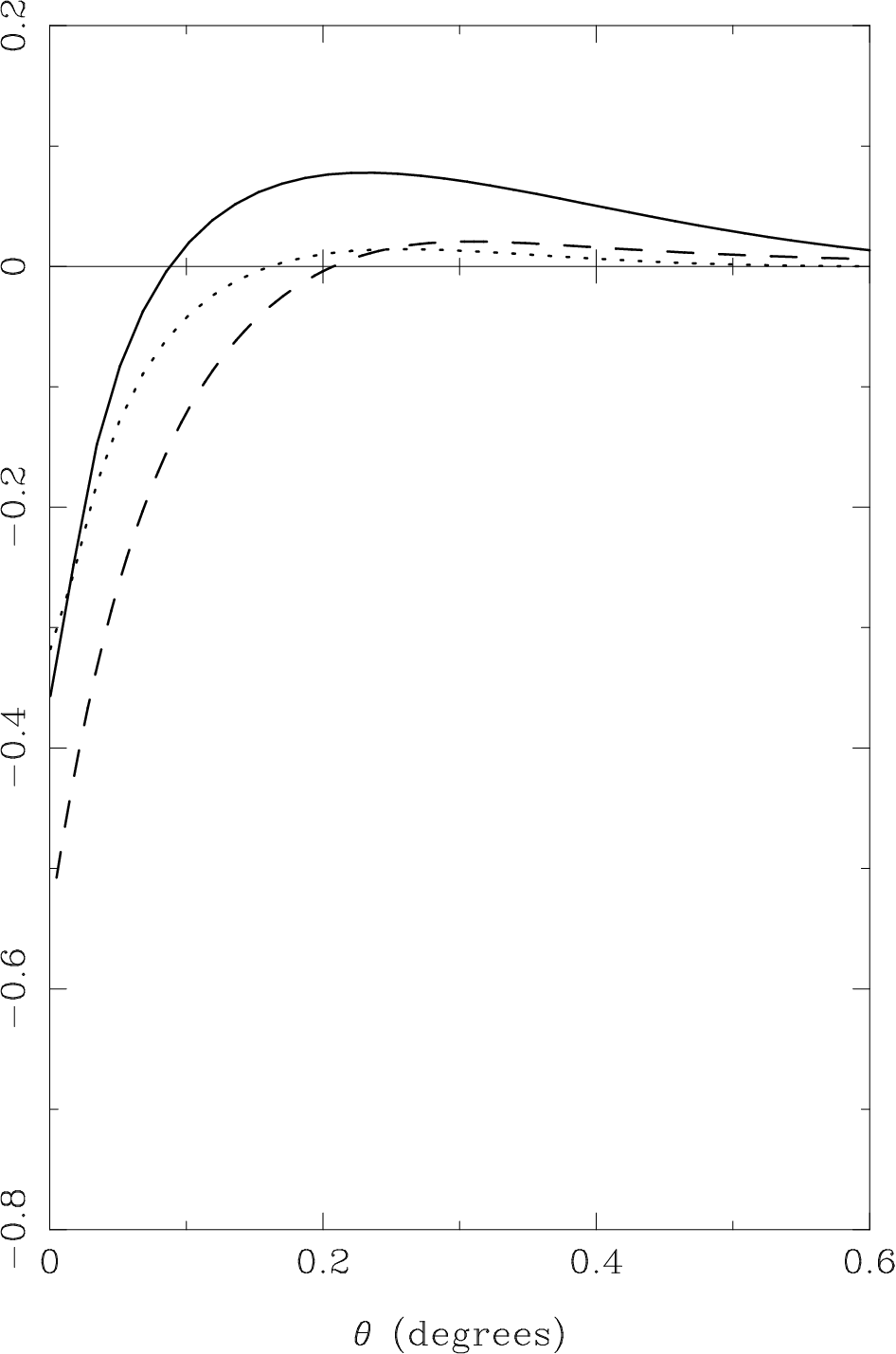, width=5.1cm, height=8.2cm}
}
\caption{Estimates of the angular distribution of the CMB anisotropy due
to the presence of a collapsing cluster on the line of sight. From
left to right, the three figures are for clusters at redshift $z=1$,
$z=2$ and $z=3$ respectively.  All clusters have the same central
density $\rho_c = 10^4 h_0^{1/2}{\rm~p~m^{-3}}$ and core radius $R_c =
0.23 h_0^{-1}$~Mpc.  The anisotropy is in units of $\Delta T/T$ as a
function of the projected angle $\theta$ on the sky from the centre of
the cluster. The solid, dashed and dotted lines are for the \OPEN,
\FLATL and \FLAT cosmological models respectively.  Note that the
angular scale is different for each figure.}
\label{fig:dt_t}
\end{figure*}

It is also useful to estimate the temperature anisotropy as a function
of the projected angle $\theta$ on the sky from the centre of the
cluster.  Profiles of $\Delta T/T$ are given in Fig.~\ref{fig:dt_t}
for the three \FLAT, \FLATL and \OPEN models and for clusters at
$z=1$, $z=2$ and $z=3$. As noticed earlier, the magnitude of the
effect gets smaller as the redshift of the cluster
increases. Fig.~\ref{fig:dt_t} reveals that the angular distribution
of the anisotropy is strongly dependent upon the cosmological model.
In particular for the case where the cluster is located at $z=1$, the
\OPEN model estimate predicts a large and extended temperature
{\em increment} on a scale of about one degree.

The fact that the temperature anisotropy can take positive as well as
negative values is not surprising when looking at
equation~(\ref{eq:dt_t}).  As long as $\theta$ is not too small
$(-\partial \Delta/\partial r)$ is negative, while $(-\Delta/r)$ is
always positive, as seen in Fig.\ref{fig:du_dr}.  Typically, there are
a negative and a positive contributions to the temperature anisotropy
$\Delta T/T$ which are weighted by $\cos^2\chi$ and $\sin^2\chi$
respectively. For larger $\theta$, $\sin^2\chi$ is bigger and the
positive exceeds the negative contribution.
\begin{figure}
\centerline{\epsfig{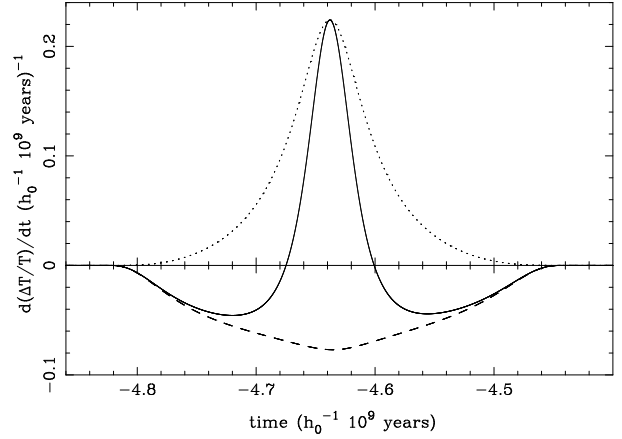}}
\caption{Solid line: The integrand of equation~(\ref{eq:dt_t}), as a
function of cosmic time $t$, along a photon path traversing a cluster
located at $z=1$ with an angle of view $\theta=0.5$~degrees from the
centre of the cluster. The time t=0 corresponds to the current epoch,
and an \OPEN model is assumed here.  The central density of the
cluster is $\rho_c = 10^4 h_0^{1/2}{\rm~p~m^{-3}}$ and its core radius
$R_c = 0.23 h_0^{-1}$~Mpc. The corresponding temperature anisotropy
$\Delta T/T$ is the area under the curve.  The dashed and dotted lines
are for $(-\partial \Delta/\partial r)$ and $(-\Delta/r)$
respectively.  }
\label{fig:du_dr}
\end{figure}
The integrand of equation~(\ref{eq:dt_t}) is plotted in
Fig.~\ref{fig:du_dr} along the path of a photon passing near the
centre of a cluster located at $z=1$. The positive and negative
contributions to the resulting $\Delta T/T$ are separated out.  In
this case, the overall perturbation is $\Delta T/T \sim +0.17\times
10^{-5}$. 
%However, this effect is so small that the dominance of the
%positive against the negative contribution is not noticeable on the
%figure.

In the case of the recently favoured \FLATL cosmological model, it is
interesting to note from Fig.\ref{fig:dt_t} that the temperature {\it
decrement} is systematically stronger and extended to larger angles
than in the other two models.

\section{Gravitational Lensing and Time Delay}
In this section we apply our model to the problem of the microwave
decrement reported towards the quasar pair PC~1643+4631~A\&B (Jones et
al. 1997). This decrement is observed at 15~GHz and lies between a
pair of quasars at redshifts $z=3.79$ and 3.83. The two quasars are
separated by 198~arcsec. If the microwave decrement is assumed to be a
thermal SZ effect caused by an intervening cluster, the minimum
central temperature decrement is estimated to be $\Delta
T/T$=$-2.1\times 10^{-4}$.  However deep optical and infrared imaging
carried with the WHT and the UKIRT (Saunders et al. 1997, Haynes et
al. 1999) as well as ROSAT X-ray observations (Kneissl, Sunyaev \&
White 1998) suggest that any intervening cluster producing the
temperature decrement should be located at a redshift as high as
$z=3$.  However, the presence of a cluster is supported by the
remarkable similarity between the quasar spectra, suggesting
gravitational lensing. Saunders et al. (1997) suggest that their $\sim 1$
per cent redshift difference can be explained in terms of quasar
intrinsic evolution over a delay between the two lightpaths of
$~\sim10^3$~years. It was found in Dabrowski et al. (1999) that the
time delays obtained in the \FLAT cosmology were too short. Here we
propose to carry out a similar investigation but in the
\FLATL and \OPEN cosmological models.

In this section, we assume $H_0 = 50~{\rm km~s^{-1}Mpc^{-1}}$; the
maximum baryonic density along the photon path is still fixed to
$10^4~{\rm p~m^{-3}}$; the total gravitational mass over baryonic mass
ratio is 10. Following Dabrowski et al. (1999), we simplify the
situation by considering a source directly behind the centre of the
cluster. This source of photons is co-moving with the Hubble flow and
is located at a redshift $z=3.8$. The cluster, located at $z=1$, $z=2$
or $z=3$ is the gravitational lens and, in order to be consistent with
Jones et al. (1997) we require the Einstein ring of the lens system to
be 100~arcsec. Such a constraint allows us to fix the core radius
$R_c$ of the cluster, as given in Table~\ref{tab:lens}. The estimated
time delays $\Delta t$ are between a photon from the source travelling
straight through the centre of the cluster and one which follows a
lensed path, appearing at the Einstein ring radius. Results are given
in Table~\ref{tab:lens} as a function of the cosmological model
assumed and the redshift of the cluster.
%
%\begin{table}
%\caption{}
%\begin{center}
%\begin{tabular}{lcccc} \hline
%Model  & $z$ & $R_c$ & $\Delta t$ & $\Delta T$ \\ \hline
%\FLAT  &  1  & 0.45 & 145 & 174 \\
%\FLATL &  1  & 0.40 & 250 & \\
%\OPEN  &  1  & 0.44 & 195 & \\
%\hline
%\FLAT  &  2  & 0.75 & 205 & 611 \\
%\FLATL &  2  & 0.60 & 550 & \\
%\OPEN  &  2  & 0.70 & 425 & \\
%\hline
%\FLAT  &  3  & 2.68 & 25 & 19000\\
%\FLATL &  3  & 1.36 & & \\
%\OPEN  &  3  & 1.89 & 140 & \\
%\hline
%\end{tabular}
%\end{center}
%\label{tab:lens}
%\end{table}
%
%
\begin{table}
\caption{Gravitational lensing time delays in the frame of
the emitting source located at $z=3.8$. The lens is placed at a
reshift $z=1$, $z=2$ or $z=3$. $R_c$ is for the core radius of the
cluster quoted in Mpc and $\Delta t$ is the time delay in years. In
all cases the cluster has a central density of $\rho_c = 10^4
h_0^{1/2}{\rm~p~m^{-3}}$ at the time photons crosses it and the lens
system is characterised by an Einstein ring of 100~arcsec.  }
\begin{center}
\begin{tabular}{l|cc|cc|cc} \hline
& \multicolumn{2}{c|}{$z=1$} & \multicolumn{2}{c|}{$z=2$}& \multicolumn{2}{c|}{$z=3$}\\
\hline
Model & $R_c$ & $\Delta t$ & $R_c$ & $\Delta t$ & $R_c$ & $\Delta t$ \\
%      & Mpc & yrs &  Mpc & yrs & Mpc & yrs \\
\hline
\FLAT & 0.45 & 145 & 0.75 & 205 & 2.68 & 25 \\
\FLATL & 0.40 & 250 & 0.60 & 550 & 1.36 & 380 \\
\OPEN & 0.44 & 195 & 0.70 & 425 & 1.89 & 140 \\
\hline
\end{tabular}
\end{center}
\label{tab:lens}
\end{table}
Since the source is at fixed redshift $z=3.8$ the lensing effect has
to be stronger for more distant clusters in order to keep an Einstein
ring of 100~arcsec. This is why the core radii are getting larger with
redshift, as seen of Table~\ref{tab:lens}.  Consequently, the time
delay is increasing as well. In the \FLAT model, this is probably
leading to rather un-realistic situations such as for $z=3$ where the
core radius of the cluster is as high as $2.68$~Mpc (369~arcsec). In
this case, the light path passing through the Einstein ring at
100~arcsec is also travelling through the core of the cluster.  This
is why the estimated time delay of 25 years is so small.  We find that
in the \OPEN and particularly in the \FLATL cosmological models, the
predicted time delays are larger than for the
\FLAT case and may be enough to explain the observations related to
PC1643+4631~A\&B, if they are indeed a gravitational lens system.

\section{Conclusion}
We extend a general relativistic model for the formation of non-linear
cosmic structures (Lasenby et al. 1999; Dabrowski et al. 1999) to the
case of open cosmologies and non-zero cosmological constant. The
equations of fluid dynamics are presented as if derived from first
principles.  This is possible when working in the context of the
Newtonian gauge (Lasenby et al. 1998). The fluid streamlines are
derived analytically in the case of a flat universe with non-zero
cosmological constant.  However, in non-homogeneous regions the
streamlines have to be computed numerically.  The initial conditions
for the fluid at the early epoch corresponding to $z=1000$ are very
simple: a finite perturbation on the fluid velocity is imposed, and
the corresponding density perturbation is inferred assuming a
linearised model and that the perturbation arose from primordial
fluctuations. As expected, in order to give rise to similar structures
at a later time, we find that the initial perturbation has to be
larger in the case of a low density universe than in a critical
density universe.

Photon geodesics can be integrated numerically (see Lasenby et
al. 1999) and we study the gravitational effect of the collapsing
galaxy clusters on CMB photons. Clusters at redshifts of $z=1$, $z=2$
and $z=3$ are considered. We find that the central temperature
decrement $\Delta T/T$ is systematically slightly stronger and
extended to larger angles in the non-zero cosmological constant
case. It was also found in the open cosmology model that the CMB
anisotropy could reach relatively large positive values.

Finally, we apply our model to the microwave decrement reported
towards the quasar pair PC~1643+4631~A\&B (Jones et al. 1997).  Here
we are particularly interested in estimating the time delay in the
source frame between the two quasar images. We find a time delay of
$\sim 550$ years in the case of a flat universe with non-zero
cosmological constant ($\Omega_0=0.3$, $\Omega_{\Lambda}=0.7$). If the
quasar images are indeed lensed images from a single source, this time
delay may be sufficiently large to explain the 1 per cent difference
in the quasar redshifts.

\section*{Acknowledgements}

YD would like to thank Trinity Hall, Cambridge for support in the form
of a Research Fellowship and John Peacock for valuable comments. We
also thank Anthony Challinor for many useful discussions.

\bsp  
\label{lastpage}
\end{document}